\documentclass[10pt, conference]{IEEEtran}
\usepackage{subfigure}
\usepackage{setspace}
\usepackage{amsmath}
\usepackage{amssymb}
\usepackage{amsfonts}
\usepackage{amscd}
\usepackage{mathrsfs}
\usepackage[final]{graphicx}
\usepackage{graphicx}
\usepackage{psfrag}
\usepackage{color}
\usepackage{url}
\usepackage{bm}

\begin{document}

\title{Orthogonal vs Non-Orthogonal Multiple Access with Finite Input Alphabet and Finite Bandwidth}

\author{
\authorblockN{J. Harshan}
\authorblockA{Dept. of ECE, Indian Institute of Science \\
Bangalore 560012, India\\
Email:harshan@ece.iisc.ernet.in\\
}
\and
\authorblockN{B. Sundar Rajan}
\authorblockA{Dept. of ECE, Indian Institute of Science \\
Bangalore 560012, India\\
Email:bsrajan@ece.iisc.ernet.in\\
}
}

\maketitle

\begin{abstract}
For a two-user Gaussian multiple access channel (GMAC), frequency division multiple access (FDMA), a well known orthogonal-multiple-access (O-MA) scheme has been preferred to non-orthogonal-multiple-access (NO-MA) schemes since FDMA can achieve the sum-capacity of the channel with only single-user decoding complexity [\emph{Chapter 14, Elements of Information Theory by Cover and Thomas}]. However, with finite alphabets, in this paper, we show that NO-MA is better than O-MA for a two-user GMAC. We plot the constellation constrained (CC) capacity regions of a two-user GMAC with FDMA and time division multiple access (TDMA) and compare them with the CC capacity regions with trellis coded multiple access (TCMA), a recently introduced NO-MA scheme. Unlike the Gaussian alphabets case, it is shown that the CC capacity region with FDMA is strictly contained inside the CC capacity region with TCMA. In particular, for a given bandwidth, the gap between the CC capacity regions with TCMA and FDMA is shown to increase with the increase in the average power constraint. Also, for a given power constraint, the gap between the CC capacity regions with TCMA and FDMA is shown to decrease with the increase in the bandwidth. Hence, for finite alphabets, a NO-MA scheme such as TCMA is better than the well known O-MAC schemes, FDMA and TDMA which makes NO-MA schemes worth pursuing in practice for a two-user GMAC.

\end{abstract}


\section{Introduction}
\label{sec1}
\begin{figure}
\centering
\includegraphics[width=3in]{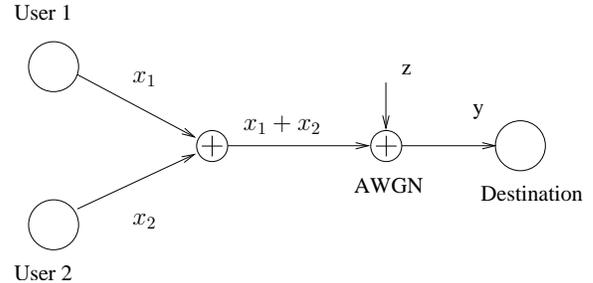}
\caption{Two-user Gaussian MAC model} 
\label{gmac_model_fig}
\end{figure}

\begin{figure*}
\centering
\includegraphics[width=4.2in]{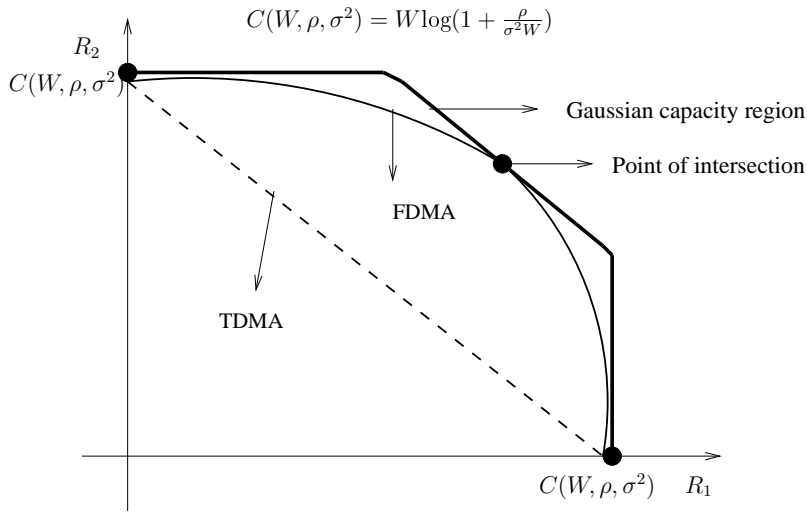}
\caption{Achievable rate pairs (in bits per second) for TDMA and FDMA for a total bandwidth of $W$ Hertz} 
\label{Capacity_region_tdma_fdma}
\end{figure*}

Works on coding for Gaussian multiple access channel (GMAC) with finite input alphabets have been reported in \cite{FTL}-\cite{WCA}. Recently, constellation constrained (CC) capacity regions \cite{Eb} of a two-user GMAC (shown in Fig. \ref{gmac_model_fig}) have been computed in \cite{HaR1} wherein the 
 angles of rotation between the alphabets of the users which enlarge the CC capacity regions have also been computed. For this channel model, code pairs based on trellis coded modulation (TCM) \cite{Ub} are proposed in \cite{HaR2}-\cite{HaR4} such that \emph{any rate pair} within the CC capacity region can be approached. Such a multiple access scheme which employs capacity approaching trellis codes is referred as trellis coded multiple access (TCMA). In this paper, multiple access schemes are referred to as non-orthogonal multiple access (NO-MA) schemes whenever the two users transmit simultaneously during the same time and in the same frequency band. TCMA is an example for a NO-MA scheme. Henceforth, throughout the paper, \textit{CC capacity regions obtained in \cite{HaR1} are referred to as CC capacity regions with TCMA since TCMA is shown to achieve any rate pair on the CC capacity region} \cite{HaR4}.\\
\indent For a two-user GMAC with Gaussian distributed continuous input alphabets, it is well known that successive interference cancellation decoder can achieve any point on the capacity region, provided the codebooks contain infinite length codewords \cite{ThC, RiU}. It is also known that time division multiple access (TDMA) and frequency division multiple access (FDMA), two of the widely known orthogonal multiple access (O-MA) techniques do not achieve all the points on the capacity region. In particular, if FDMA is used such that the bandwidth allocated to each user is proportional to its transmit power, then one of the points on the capacity region can be achieved. Note that such an achievable point lies on the sum-capacity line segment of the capacity region \cite{ThC} (note that apart from the point on the sum capacity line, FDMA also intersects the capacity region at the axes points). The set of rate pairs that can be achieved using TDMA and FDMA are provided in Fig. \ref{Capacity_region_tdma_fdma} along with the capacity region wherein the total bandwidth (for both the users) is $W$ Hertz, the power constraint for each user is $\rho$ Watts and the variance of the AWGN is $\sigma^{2}$.\\
\indent In this paper, we compute the CC capacity regions of a two-user GMAC with the well known O-MA schemes such as TDMA and FDMA. On the similar lines of the study for Gaussian input alphabets (as shown in Fig. \ref{Capacity_region_tdma_fdma}), we study the differences among the CC capacity regions of a two-user GMAC with TDMA, FDMA and TCMA. Throughout the paper, the terms alphabet and signal set are used interchangeably. The main contributions of this paper may be summarized as below:
\begin{itemize}

\item We compute the CC capacity regions of a two-user GMAC when O-MA schemes such as TDMA and FDMA are employed for finite bandwidth.  
\item Since FDMA with Gaussian alphabets achieve one of the sum-capacity points with single-user decoding complexity, the authors of \cite{ThC} have stated that \emph{"the improvement in the capacity due to multiple access ideas such as the one achieved by the successive interference decoder (a NO-MA scheme) may not be sufficient to warrant the increased complexity"} (see page 407, Section 14.3.6 of Chapter 14 in \cite{ThC}). In this paper, we point out that the above comment in \cite{ThC} is not valid for a GMAC with finite alphabets which is the case in practical scenarios. In particular, we show that NO-MA schemes such as TCMA can provide substantial improvement in the capacity when compared to O-MA schemes such as FDMA and TDMA for finite alphabets which in-turn makes TCMA worth pursuing in practice for a two-user GMAC. We highlight that this result is not apparent unless CC capacity regions with FDMA and TCMA are plotted (Section \ref{sec4}).
\item We also show that the gap between the CC capacity regions with TCMA and FDMA is a function of the bandwidth, $W$ Hertz and the average power constraint, $\rho$ Watts. It is shown that, (i) for a fixed $W$, the gap between the CC capacity regions with FDMA and TCMA \emph{\textbf{increases}} with the increase in $\rho$ (see Fig. \ref{region_2db}, Fig. \ref{region_5db} and Fig. \ref{region_8db} for a fixed $W$ and varying $\rho$), and (ii) for a fixed $\rho$, the gap between the CC capacity regions with FDMA and TCMA \emph{\textbf{decreases}} with the increase in $W$ (see Fig. \ref{region_2db} and  Fig. \ref{region_2db_W_5} for a fixed $\rho$ and varying $W$) (Section \ref{sec4_subsec1}).
\item On the similar lines of the results for Gaussian alphabets, the CC capacity region with TDMA is shown to be well contained inside the CC capacity region with TCMA (Section \ref{sec4_subsec2}).
\end{itemize}

\indent The result presented in this paper is another example to illustrate the differences in the system behaviour in terms of capacity when the input alphabets are constrained to have finite cardinality. An earlier example can be found in \cite{HaR1}, wherein a relative angle of rotation between the input alphabets is shown to enlarge the CC capacity region with TCMA \footnote{Note that the CC capacity region for a two-user GMAC \cite{HaR1} is referred as the CC capacity region with TCMA since TCMA is shown to achieve any rate pair on the CC capacity region \cite{HaR3}}. Note that such a capacity advantage is not applicable for Gaussian alphabets. For example, see Fig. \ref{CC_region_equal}, which depicts the advantage of rotations on the CC capacity region with TCMA for QPSK signal sets. For more details on the advantages of rotation on the CC capacity region, we refer the readers to \cite{HaR3}.

\textit{Notations:} Cardinality of a set $\mathcal{S}$ is denoted by $|\mathcal{S}|$. Absolute value of a complex number $x$ is denoted by $|x|$ and $E \left[x\right]$ denotes the expectation of the random variable $x$. A circularly symmetric complex Gaussian random vector, $\textbf{x}$ with mean $\bm{\mu}$ and covariance matrix $\mathbf{\Gamma}$ is denoted by $\textbf{x} \sim \mathcal{CN} \left(\bm{\mu}, \mathbf{\Gamma} \right)$.

The remaining content of the paper is organized as follows: In Section \ref{sec2}, the two-user GMAC model considered in this paper is presented wherein we recall the CC capacity regions with TCMA. In Section \ref{sec3}, we revisit the achievable rate pairs for FDMA and TDMA with Gaussian alphabets. In Section \ref{sec4}, we compute the CC capacity regions with FDMA and TDMA and show the optimality of TCMA over FDMA and TDMA. Section \ref{sec5} constitutes conclusion and some directions for possible future work.

\begin{figure*}
{\small
\begin{equation}
\label{mi}
I(\sqrt{\frac{\rho}{W}}x_{2} : y) = \mbox{log}_{2}(N_{2}) - \frac{1}{N_{1}N_{2}}\sum_{k_{1} = 0}^{N_{1} - 1}\sum_{k_{2} = 0}^{N_{2} - 1}E\left[\mbox{log}_{2}\left[ \frac{\sum_{i_{1} = 0}^{N_{1} - 1}\sum_{i_{2} = 0}^{N_{2} - 1} \mbox{exp}\left(- |\sqrt{\frac{\rho}{W}}\left(x_{1}(k_{1}) + x_{2}(k_{2}) - x_{1}(i_{1}) - x_{2}(i_{2})\right)  + z|^{2}/\sigma^{2}\right)}{\sum_{i_{1} = 0}^{N_{1} - 1} \mbox{exp}\left(- |\sqrt{\frac{\rho}{W}}x_{1}(k_{1}) - \sqrt{\frac{\rho}{W}}x_{1}(i_{1})+ z|^{2}/\sigma^{2}\right)}\right] \right].
\end{equation}
}

\begin{equation}
\label{mi1}
I(\sqrt{\frac{\rho}{W}} x_{1} : y ~|~x_{2} ) = \mbox{log}_{2}(N_{1}) - \frac{1}{N_{1}}\sum_{k_{1} = 0}^{N_{1} - 1}E\left[\mbox{log}_{2}\left[ \frac{\sum_{i_{1} = 0}^{N_{1} - 1} \mbox{exp}\left(- |\sqrt{\frac{\rho}{W}}x_{1}(k_{1}) -  \sqrt{\frac{\rho}{W}}x_{1}(i_{1}) + z|^{2}/\sigma^{2}\right)}{\mbox{exp}\left(- |z|^{2}/\sigma^{2}\right)}\right] \right].
\end{equation}
\hrule
\end{figure*}

\section{Channel model and CC capacity regions of GMAC}
\label{sec2}
The model of a two-user Gaussian MAC as shown in Fig. \ref{gmac_model_fig} consists of two users that need to convey information to a single destination. It is assumed that User-1 and User-2 communicate to the destination at the same time and in the same frequency band of $W$ Hertz. Symbol level synchronization is assumed at the destination. The two users are equipped with alphabets $\mathcal{S}_{1}$ and $\mathcal{S}_{2}$ of size $N_{1}$ and $N_{2}$ respectively such that for $x_{i} \in \mathcal{S}_{i}$, we have  $E[|x_{i}|^{2}] = 1$. In other words, the signal sets $\mathcal{S}_{1}$ and $\mathcal{S}_{2}$ have unit average energy. When User-1 and User-2 transmit symbols $x_{1}$ and $x_{2}$ simultaneously, the destination receives a symbol $y$ given by,
\begin{equation}
\label{gmac_model}
y = \sqrt{\frac{\rho}{W}} x_{1} + \sqrt{\frac{\rho}{W}} x_{2} + z ~~ \mbox{where }  z \sim \mathcal{CN} \left(0, \sigma^{2} \right)
\end{equation}
where $\rho$ is the average power constraint for the two users. Throughout the paper, we assume equal average power constraint for the two users. Extension of the results presented in this paper to the unequal power case is straight forward. We assume that every channel use consumes $T$ seconds for each user (where we assume $\frac{1}{T} = W$ Hertz) and hence $\frac{\rho}{W}$ becomes the average energy constraint for the two users.\\
\begin{figure}
\centering
\includegraphics[width=3.4in]{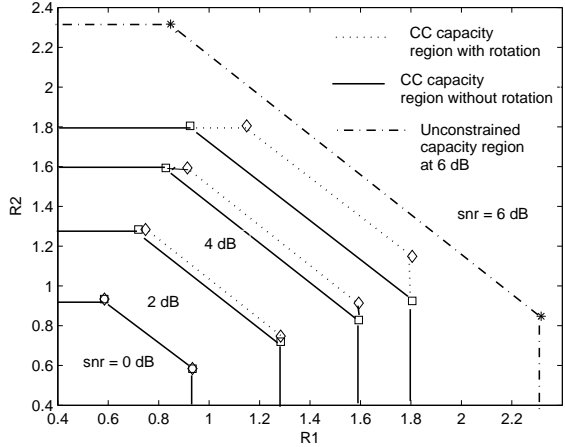}
\caption{CC capacity regions with TCMA of QPSK signal sets with and without rotation for two-user GMAC where snr = $\frac{\rho}{W}$ and $\sigma^{2} = 1.$} 
\label{CC_region_equal}
\end{figure}
\indent In \cite{HaR1}, CC capacity regions for two-user Gaussian multiple access channels have been computed for some well known alphabets such as $M$-PSK, $M$-QAM etc. Applying the results of \cite{HaR1} to the channel model in \eqref{gmac_model}, the set of CC capacity values (in bits per channel use) that define the boundary of the CC capacity region are as given below,
$$R_{1} ~\leq ~ I(\sqrt{\frac{\rho}{W}} x_{1} : y ~|~ x_{2}),$$
$$R_{2} ~\leq ~ I(\sqrt{\frac{\rho}{W}} x_{2} : y ~|~ x_{1}) \mbox{ and }$$
\begin{equation}
\label{capacity_region_bpc}
R_{1}  + R_{2} ~ \leq ~ I(\sqrt{\frac{\rho}{W}} x_{1} + \sqrt{\frac{\rho}{W}} x_{2} : y ),
\end{equation}
where the expressions for $I(\sqrt{\frac{\rho}{W}} x_{2}: y)$,  $I(\sqrt{\frac{\rho}{W}} x_{1} : y ~|~ x_{2})$ are shown in \eqref{mi} and \eqref{mi1} at the top of this page. The term $I(\sqrt{\frac{\rho}{W}} x_{1} + \sqrt{\frac{\rho}{W}} x_{2}: y)$ can be calculated as $I(\sqrt{\frac{\rho}{W}} x_{2}: y)$ + $I(\sqrt{\frac{\rho}{W}} x_{1} : y ~|~ x_{2})$. Examples of CC capacity regions are shown in Fig. \ref{CC_region_equal} for QPSK alphabets. Since every channel use consumes $T$ seconds, the CC capacity pairs (in bits per seconds) that define the CC capacity region are given by
$$R_{1} ~\leq ~ WI(\sqrt{\frac{\rho}{W}} x_{1} : y ~|~ x_{2}),$$
$$R_{2} ~\leq ~ WI(\sqrt{\frac{\rho}{W}} x_{2} : y ~|~ x_{1}) \mbox{ and }$$
\begin{equation}
\label{capacity_region}
R_{1}  + R_{2} ~ \leq ~ WI(\sqrt{\frac{\rho}{W}} x_{1} + \sqrt{\frac{\rho}{W}} x_{2} : y ).
\end{equation}
 
\indent In the following section, we will revisit the capacity region of a two-user Gaussian MAC (referred as the unconstrained capacity region of two-user GMAC) and recall the set of achievable rate pairs when FDMA and TDMA are employed. 

\section{Unconstrained capacity regions of two-user GMAC}
\label{sec3}
For computing the CC capacity regions, the input alphabets are assumed to take values with uniform distribution \cite{HaR1}. However, if the input alphabets are continuous and distributed as $x_{1}, x_{2} \sim \mathcal{CN} \left(0,  1\right)$, then the received symbol, $y$ is Gaussian distributed as $\mathcal{CN} \left(0,  \frac{2\rho}{W} + \sigma^{2} \right)$. Considering the bandwidth of W Hertz, the unconstrained capacity region in bits per second is given by 
$$R_{1} \leq W\mbox{log}_{2}(1 + \frac{\rho}{\sigma^{2} W}) = C(W, \rho, \sigma^{2}),$$
$$R_{2} \leq W\mbox{log}_{2}(1 + \frac{\rho}{\sigma^{2} W}) = C(W, \rho, \sigma^{2}) \mbox{ and }$$
\begin{equation}
\label{sum_rate_gaussian_region}
R_{1} + R_{2}  \leq W\mbox{log}_{2}(1 + \frac{2\rho}{\sigma^{2} W}).
\end{equation}
\noindent From the results of \cite{ThC, RiU}, it is well known that any rate pair within the capacity region can be achieved by using successive interference cancellation decoder provided the codebooks have infinite length codewords. 
When the two users employ FDMA, let $W_{1} = \alpha W$ and $W_{2} = (1-\alpha)W$ be the bandwidth occupied by User-1 and User-2 respectively where $0 < \alpha < 1$. For such a scheme, 
the maximum achievable rate pairs (in bits per second) for the two users are given by 
\begin{equation}
\label{fdma_gaussian _rates}
R_{1} \leq W_{1}\mbox{log}_{2}(1 + \frac{\rho}{\sigma^{2} W_{1}}) \mbox{ and } R_{2} \leq W_{2}\mbox{log}_{2}(1 + \frac{\rho}{\sigma^{2} W_{2}}).
\end{equation}
The maximum achievable sum rate when $\alpha = 0.5$ is given by,
\begin{equation}
\label{sum_rate_FDMA_gaussian}
R_{1} + R_{2} \leq W\mbox{log}_{2}(1 + \frac{2\rho}{\sigma^{2} W})
\end{equation}
which is equal to the sum capacity of the two-user Gaussian MAC given in \eqref{sum_rate_gaussian_region}. Therefore, for Gaussian alphabets, FDMA can achieve one of the points on the maximum sum rate line of the unconstrained capacity region. Proceeding similarly for the case of TDMA, one obtains the set of achievable rate pairs as shown in Fig. \ref{Capacity_region_tdma_fdma}.\\
\indent On the similar lines of the discussion in this section, in the following section, we will obtain the CC capacity regions with FDMA and TDMA.
\section{CC Capacity regions with FDMA and TDMA}
\label{sec4}
\begin{figure}
\centering
\includegraphics[width=3.6in]{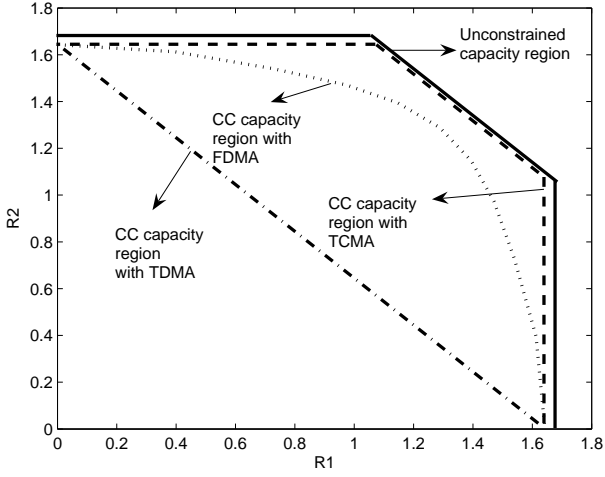}
\caption{CC Capacity regions (in bits/sec) with FDMA, TDMA and TCMA for two-user GMAC with QPSK signal sets when $\rho$ = 2 dB, $\sigma^{2}$ = 1, and $W$ = 2 Hertz.} 
\label{region_2db}
\end{figure}\
\begin{figure}
\centering
\includegraphics[width=3.6in]{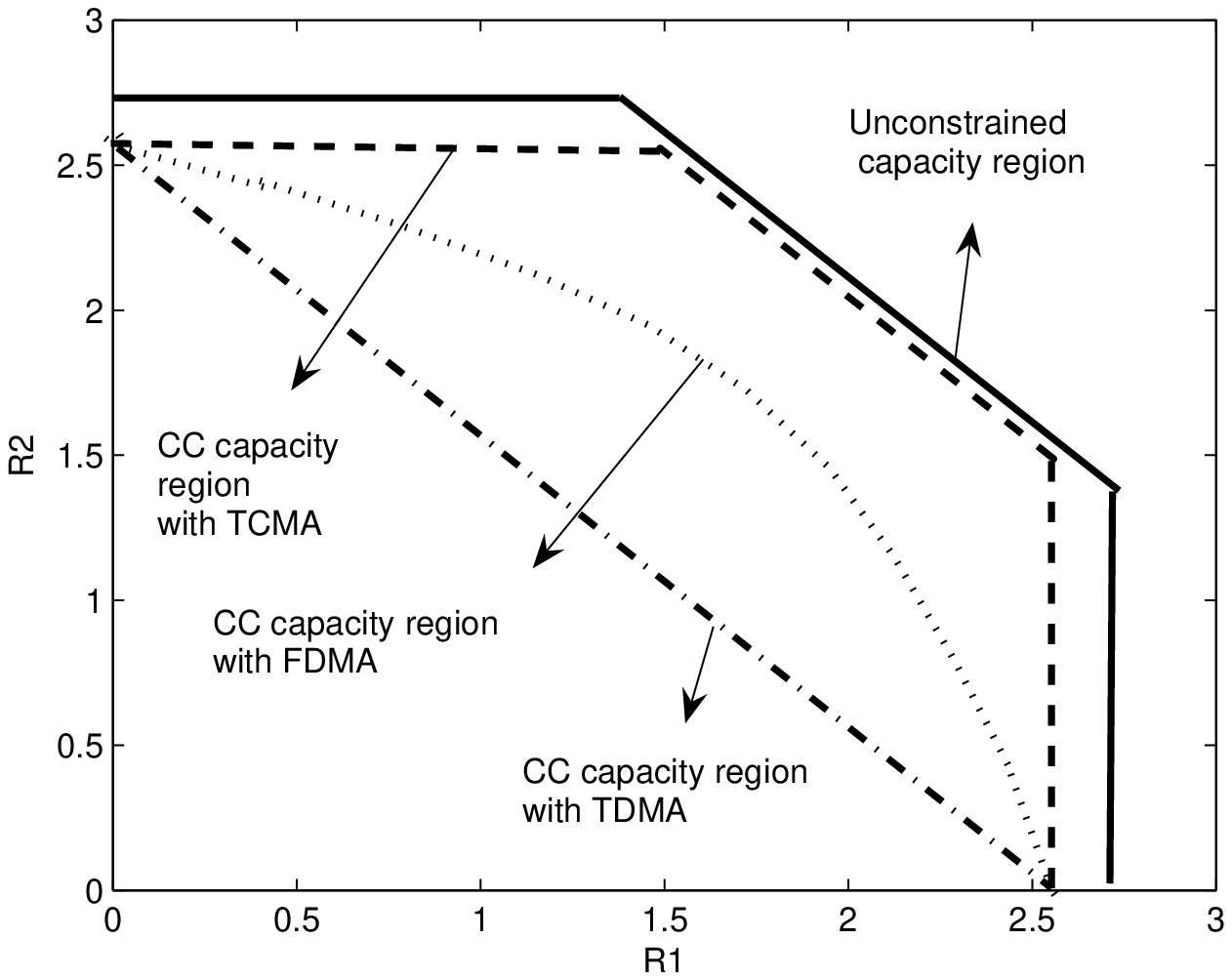}
\caption{CC Capacity regions (in bits/sec) with FDMA, TDMA and TCMA for two-user GMAC with QPSK signal sets when $\rho$ = 5 dB, $\sigma^{2}$ = 1, and $W$ = 2 Hertz.} 
\label{region_5db}
\end{figure}
\begin{figure}
\centering
\includegraphics[width=3.6in]{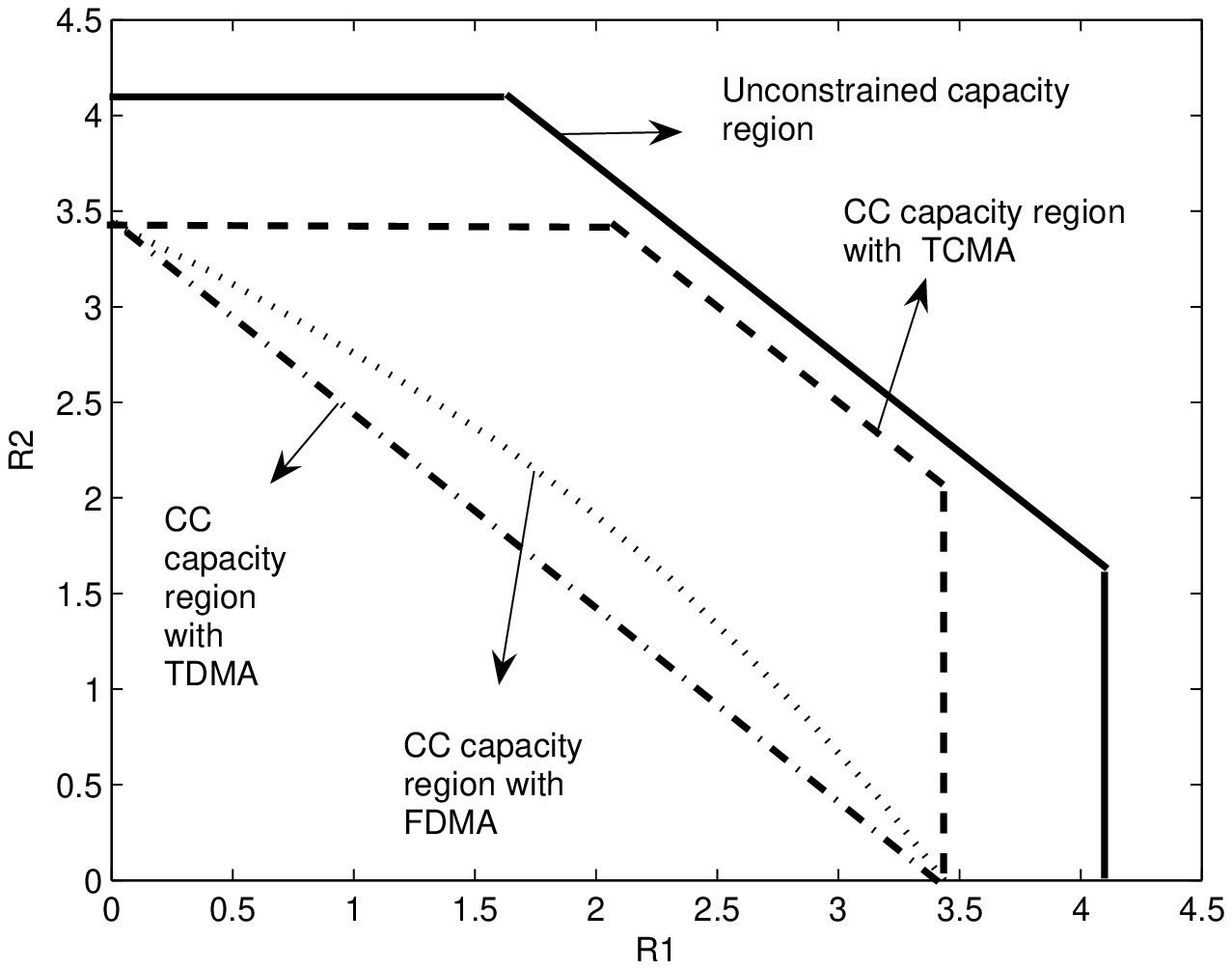}
\caption{CC Capacity regions (in bits/sec) with FDMA, TDMA and TCMA for two-user GMAC with QPSK signal sets when $\rho$ = 8 dB, $\sigma^{2}$ = 1, and $W$ = 2 Hertz.} 
\label{region_8db}
\end{figure}
\begin{figure}
\centering
\includegraphics[width=3.6in]{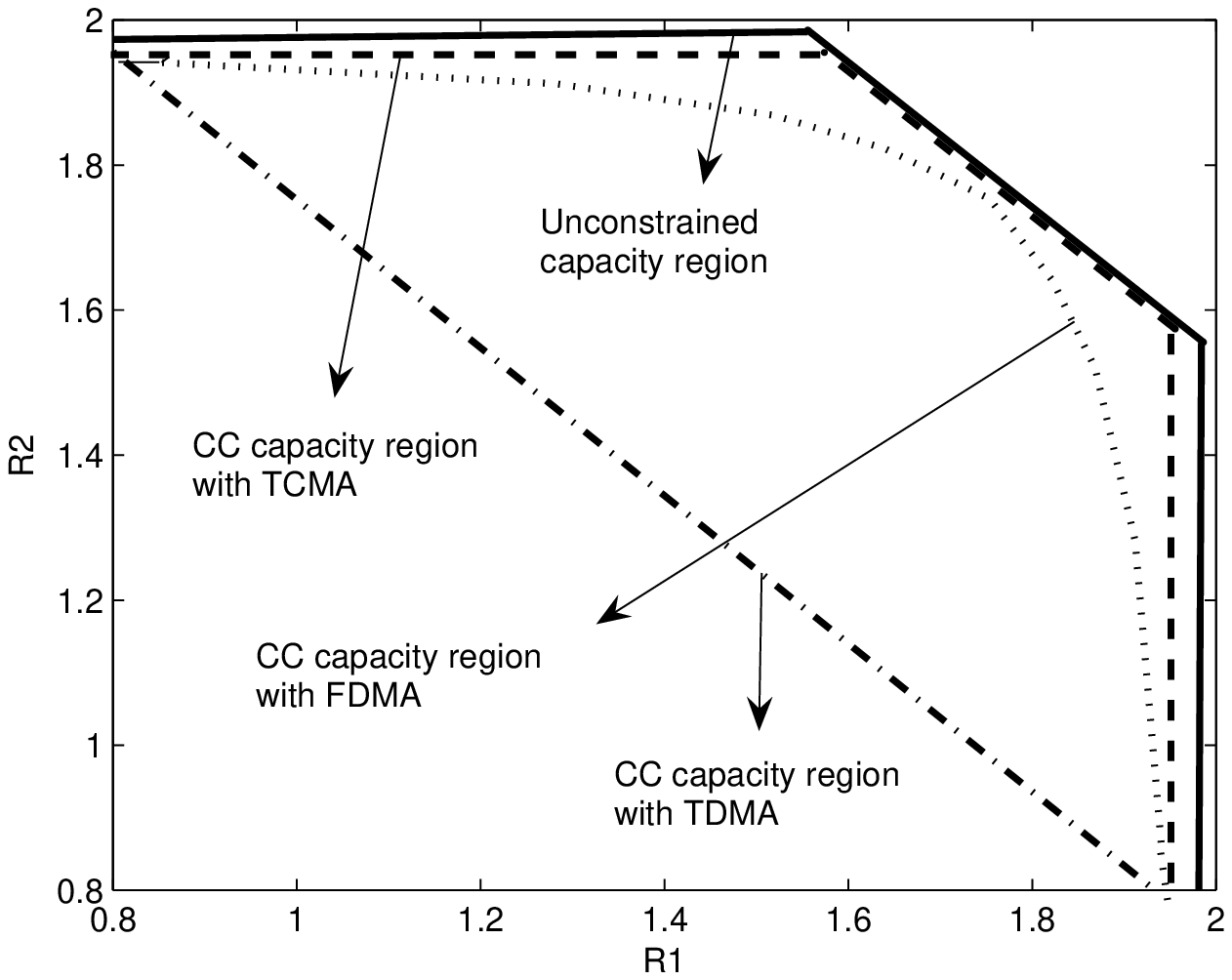}
\caption{CC Capacity regions (in bits/sec) with FDMA, TDMA and TCMA for two-user GMAC with QPSK signal sets when $\rho$ = 2 dB, $\sigma^{2}$ = 1, and $W$ = 5 Hertz.} 
\label{region_2db_W_5}
\end{figure}
\begin{figure}
\centering
\includegraphics[width=3.6in]{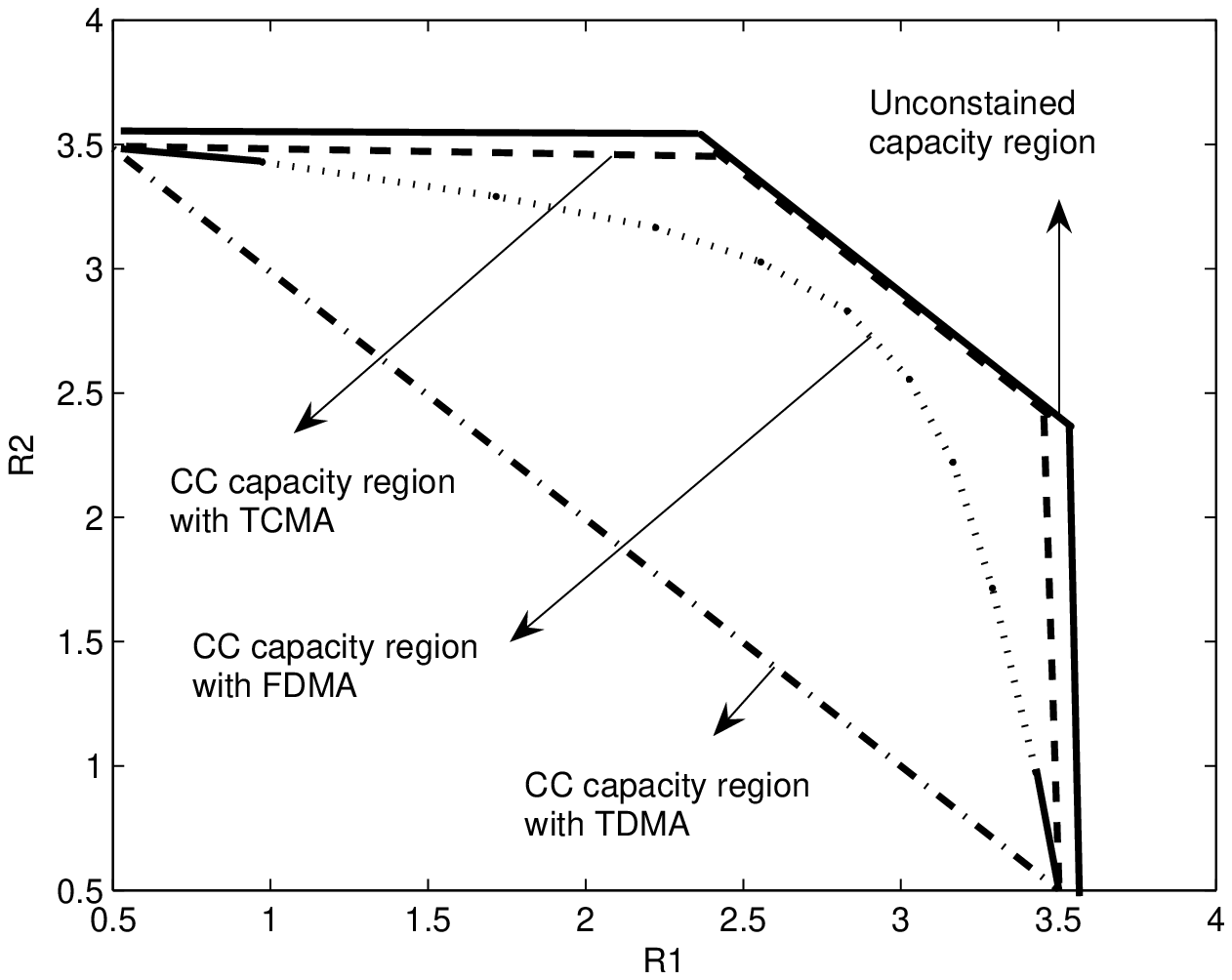}
\caption{CC Capacity regions (in bits/sec) with FDMA, TDMA and TCMA for two-user GMAC with QPSK signal sets when $\rho$ = 5 dB, $\sigma^{2}$ = 1, and $W$ = 5 Hertz.} 
\label{region_5db_W_5}
\end{figure}
\begin{figure}
\centering
\includegraphics[width=3.6in]{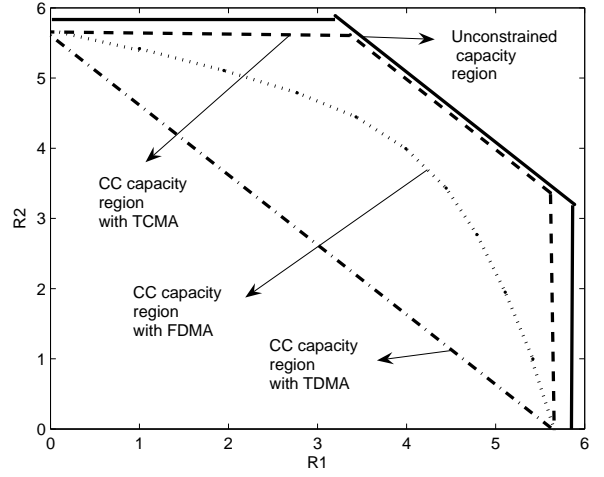}
\caption{CC Capacity regions (in bits/sec) with FDMA, TDMA and TCMA for two-user GMAC with QPSK signal sets when $\rho$ = 8 dB, $\sigma^{2}$ = 1, and $W$ = 5 Hertz.} 
\label{region_8db_W_5}
\end{figure}
In this section, we first compute the CC capacity regions with FDMA. Let $W_{1} = \alpha W$ and $W_{2} = (1-\alpha)W$ be the disjoint band of frequencies occupied by User-1 and User-2 respectively where $0 < \alpha < 1$. Hence, for each $i = 1, 2$, User-$i$ views a Single-Input Single-Output (SISO) AWGN channel to the destination with the input alphabet $\mathcal{S}_{i}$, bandwidth $W_{i}$ and energy constraint, $\frac{\rho}{W_{i}}$. The CC capacity values (in bits per second) for the two users are given by,
\begin{equation*}
~~~~R_{1} \leq W_{1}I\left(\sqrt{\frac{\rho}{W_{1}}} x_{1} : y ~|~ x_{2}\right) \mbox{ and }
\end{equation*}
\begin{equation*}
 R_{2} \leq W_{2}I\left(\sqrt{\frac{\rho}{W_{2}}} x_{2} : y ~|~ x_{1}\right) .
\end{equation*}
Note that when $\alpha = 0.5$, the CC sum capacity with FDMA is given by,

{\small
\begin{equation*}
R_{1}  + R_{2} \leq \frac{W}{2}I\left(\sqrt{\frac{2\rho}{W}} x_{1} : y ~|~ x_{2}\right) + \frac{W}{2}I\left(\sqrt{\frac{2\rho}{W}} x_{2} : y ~|~ x_{1}\right).
\end{equation*}
}

\noindent If we assume the two users to employ identical signal sets, then the CC sum capacity (denoted by $R_{1} + R_{2} | _{\mbox{FDMA}, \alpha = 0.5})$ is given by
\begin{equation}
\label{sum_rate_FDMA_finite}
R_{1} + R_{2} | _{\mbox{FDMA}, \alpha = 0.5} \leq WI\left( \sqrt{\frac{2\rho}{W\sigma^{2}}} x_{1} : y ~| ~x_{2}\right)\\
\end{equation}
wherein without loss of generality, we have used the variable $x_{1}$ for both the users. 
\subsection{Comparing the CC capacity regions of TCMA and FDMA}
\label{sec4_subsec1}
The CC sum capacity with TCMA (denoted by $R_{1}  + R_{2} | _{\mbox{TCMA}}$) given in \eqref{capacity_region} is 
\begin{equation}
\label{cc_sum_capacity}
R_{1}  + R_{2} | _{\mbox{TCMA}}~ \leq ~ WI\left(\sqrt{\frac{\rho}{W}} x_{1} + \sqrt{\frac{\rho}{W}} x_{2} : y \right).
\end{equation}
\noindent Comparing \eqref{sum_rate_FDMA_finite} and \eqref{cc_sum_capacity}, it is not straightforward to comment whether, the CC sum capacity offered by FDMA is equal to or away from the CC sum capacity with TCMA. Therefore, in Fig. \ref{region_2db}, Fig. \ref{region_5db} and Fig. \ref{region_8db}, we plot the CC capacity regions with TCMA and FDMA for QPSK (quadrature phase shift keying) signal sets when bandwidth, $W = 2$ Hertz. Similarly, in Fig. \ref{region_2db_W_5}, Fig. \ref{region_5db_W_5} and Fig. \ref{region_8db_W_5}, CC capacity regions with TCMA and FDMA are presented  for QPSK signal sets when bandwidth, $W = 5$ Hertz. The plots show that the CC capacity region with FDMA is strictly enclosed within the CC capacity region with TCMA. Note that, for a given value of $W$, the difference between the regions with FDMA and TCMA becomes significant with the increased value of $\rho$ (see Fig. \ref{region_2db}, Fig. \ref{region_5db} and Fig. \ref{region_8db}). In particular, the plots show the following inequality,
\begin{equation*}
R_{1} + R_{2} | _{\mbox{FDMA}, \alpha = 0.5} \leq R_{1}  + R_{2} | _{\mbox{TCMA}}.
\end{equation*}
\begin{figure}
\centering
\includegraphics[width=3.5in]{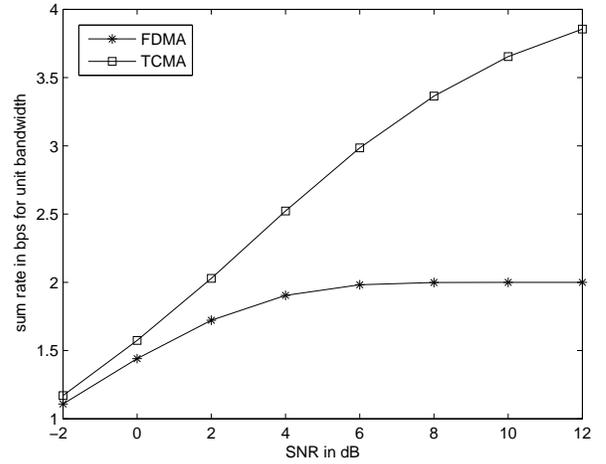}
\caption{CC sum capacity in bits per second per Hertz where SNR = $\gamma$.} 
\label{sum_rate_array}
\end{figure}
\begin{table*}
\caption{The value of $\mu$ in percentage for different values of $\gamma$}
\begin{center}
\begin{tabular}{|c|c|c|c|c|c|c|c|c|c|c|c|c|c|c|c|c|c|}
\hline $\gamma$ in dB & -2 & 0 & 2 & 4 & 6 & 8 & 10 & 12 \\
\hline $\mu$ in $\%$ & 5.46 & 9.31 & 17.79 & 32.48 & 50.55 & 68.33 & 82.70 & 92.78\\
\hline 
\end{tabular} 
\end{center}
\label{rate_table_designs_1}
\hrule
\end{table*}
\noindent Note that the difference between $R_{1} + R_{2} | _{\mbox{FDMA}, \alpha = 0.5}$ and $R_{1}  + R_{2} | _{\mbox{TCMA}}$ depends on $W$ for a given value of $\rho \mbox{ and } \sigma^{2}$. In order to make the above difference component independent of the bandwidth $W$, we calculate the percentage increase in the CC sum capacity from $R_{1} + R_{2} | _{\mbox{FDMA}, \alpha = 0.5}$ to $R_{1}  + R_{2} | _{\mbox{TCMA}}$ (denoted as $\mu$) given by,
\begin{equation*}
\mu  = \frac{WI(\sqrt{\frac{\rho}{W}} x_{1} + \sqrt{\frac{\rho}{W}} x_{2} : y ) - WI( \sqrt{\frac{2\rho}{W\sigma^{2}}} x_{1} : y ~|~ x_{2})}{WI( \sqrt{\frac{2\rho}{W}} x_{1} : y ~|~ x_{2})}
\end{equation*}
\begin{equation*}
= \frac{I(\sqrt{\frac{\rho}{W}} x_{1} + \sqrt{\frac{\rho}{W}} x_{2} : y ) - I( \sqrt{\frac{2\rho}{W}} x_{1} : y ~| ~x_{2})}{I( \sqrt{\frac{2\rho}{W}} x_{1} : y ~|~ x_{2})}.
\end{equation*}
\noindent Replacing $\gamma = \frac{\rho}{W}$, the right hand-side of the above inequality can be written as 
\begin{equation*}
\mu= \frac{I(\sqrt{\gamma} x_{1} + \sqrt{\gamma} x_{2} : y ) - I( \sqrt{2\gamma} x_{1} : y ~|~ x_{2})}{I( \sqrt{2\gamma} x_{1} : y ~|~ x_{2})}
\end{equation*}
wherein $W$ has been absorbed in the term $\gamma$. In the Table \ref{rate_table_designs_1}, we provide the values of $\mu$ for different values of $\gamma$ when $\sigma^{2} = 1$ and the input alphabets are QPSK signal sets. The values of $I(\sqrt{\gamma} x_{1} + \sqrt{\gamma} x_{2} : y )$ and $I( \sqrt{2\gamma} x_{1} : y)$ have also been plotted as a function of $\gamma$ in Fig. \ref{sum_rate_array}. The values of $\mu$ can also be plotted against $\gamma$ for larger signal sets as well.\\
\indent From Table \ref{rate_table_designs_1}, it is clear that $\mu$ increases as the value of $\gamma$ increase. An intuitive reasoning for such a behaviour is as follows: The term $I(\sqrt{\gamma} x_{1} + \sqrt{\gamma} x_{2} : y )$ is the CC capacity of a 16 point constellation (sum alphabet of two QPSK signal sets) with average energy of $2\gamma$ whereas $I( \sqrt{2\gamma} x_{1} : y)$ is the CC capacity of a 4 point constellation (QPSK signal set) with the same average energy of $2\gamma$. Note that, asymptotically (for large values of $\gamma$), $I( \sqrt{2\gamma} x_{1} : y)$ and $I(\sqrt{\gamma} x_{1} + \sqrt{\gamma} x_{2} : y )$ saturate to 2 bits and 4 bits respectively. Therefore, as $\gamma$ increases, the term $I( \sqrt{2\gamma} x_{1} : y)$ increases at a slower rate to saturate to 2 bits however, $I(\sqrt{\gamma} x_{1} + \sqrt{\gamma} x_{2} : y )$ increases at a faster rate as its saturation is only at 4 bits. A similar reasoning also holds for alphabets with arbitrary size. However, the difference in the sum capacity may differ depending on the size of the alphabets.

\subsection{CC capacity region with TDMA}
\label{sec4_subsec2}
In subsection, we obtain the CC capacity pairs when the two users employ TDMA.  If User-1 uses the channel for $\alpha$ seconds and User-2 uses the channel for $(1-\alpha)$ seconds for some $0 < \alpha < 1$, then the CC capacity values (in bits per second) for the two users are given by 
\begin{equation*}
~~~~~R_{1} \leq \alpha WI\left(\sqrt{\frac{\rho}{W}} x_{1} : y ~|~ x_{2}\right) \mbox{ and }
\end{equation*}
\begin{equation*}
 R_{2} \leq (1-\alpha)WI\left(\sqrt{\frac{\rho}{W}} x_{2} : y ~|~ x_{1}\right).
\end{equation*}
Assuming identical alphabets for the two users, the CC sum capacity with TDMA is given by,
\begin{equation*}
R_{1}  + R_{2} \leq WI\left(\sqrt{\frac{\rho}{W}} x_{1} : y ~|~ x_{2}\right).
\end{equation*}
The set of CC capacity pairs when the two users employ TDMA are shown in Fig.\ref{region_2db}, Fig.\ref{region_5db} and Fig. \ref{region_8db} which shows that TCMA is better than TDMA for finite alphabets as well.\\
\indent We highlight that, along with the substantial improvement in the capacity, low complexity trellis codes proposed for TCMA in \cite{HaR2} makes TCMA worth pursuing in practice for a two-user GMAC.

\section{Discussion}
\label{sec5}
In this paper, we compute the CC capacity regions of a two-user GMAC when multiple access schemes such as TDMA and FDMA are employed. Unlike the Gaussian alphabets case, it is shown that the CC capacity region with FDMA is strictly contained inside the CC capacity region with TCMA, essentially showing that TCMA is better than FDMA for finite alphabets. After \cite{HaR1}, the result presented in this paper is another example to illustrate the differences in the capacity behaviour when the input alphabets are constrained to have finite cardinality. Some possible directions for further work are as follows:  
\begin{itemize}
\item In this paper, we have assumed equal average power constraint for both the users. The results presented in this paper can be extended to alphabets with unequal power constraints as well
\item The presented result can also be extended to multiple access channels with more than two users.
\end{itemize}

\section*{Acknowledgment}
This work was partly supported by the DRDO-IISc Program on Advanced Research in Mathematical Engineering to B.S. Rajan.


\begin{thebibliography}{1}

\bibitem{FTL}
F. N. Brannstrom, T. M. Aulin and L. K. Rasmussen, "Constellation-Constrained capacity for Trellis code Multiple Access Systems" in the proceedings of \emph{IEEE GLOBECOM 2001}, vol. 2, San Antonio, Texas, Nov, 2001, pp. 11-15.
\bibitem{AuE}
T. Aulin and R. Espineira, "Trellis coded multiple access (TCMA)" in the proceedings of \emph{ICC '99}, Vancouver, BC, Canada, June 1999, pp. 1177-1181.
\bibitem{FAR}
Fredrik N Brannstrom, Tor M. Aulin and Lars K. Rasmussen "Iterative Multi-User Detection of Trellis Code Multiple Access using a posteriori probabilities" in the proceedings of \emph{IEEE ICC 2001}, Finland, June 2001, pp. 11-15. 
\bibitem{WCA}
Wei Zhang, C. D'Amours and A. Yongacoglu, "Trellis Coded Modulation Design form Multi-User Systems on AWGN Channels" in the proceedings of \emph{IEEE Vehicular Technology Conference 2004}, pp. 1722 - 1726. 

\bibitem{Eb}
Ezio Biglieri, "Coding for wireless channels", \emph{Springer-Verlag New York, Inc}, 2005.
\bibitem{HaR1}
J. Harshan and B. Sundar Rajan, "Finite Signal-set Capacity of Two-user Gaussian Multiple Access Channel"  in the proceedings of \emph{IEEE International Symposium on Information Theory, (ISIT 2008)}, Toronto, Canada, July 06-11, 2008. pp. 1203 - 1207.

\bibitem{Ub}
G. Ungerboeck, "Channel coding with multilevel/phase signals," \emph{IEEE Trans. Inform. Theory}, vol. 28, no. 01, 1982, pp. 55-67.
\bibitem{HaR2}
J. Harshan and B. Sundar Rajan, "Coding for Two-User Gaussian MAC with PSK and PAM Signal Sets"  in the proceedings of \emph{IEEE International Symposium on Information Theory, (ISIT 2009)}, Seoul, South Korea, June 28- July 03, 2009, pp. 1859-1863.

\bibitem{HaR3}
J. Harshan and B. Sundar Rajan, "Coding for two-user SISO and MIMO multiple access channels", available online at arXiv:0901.0168v3 [cs.IT], January 2009.

\bibitem{HaR4}
J. Harshan and B. Sundar Rajan, "Trellis coded modulation for two-user unequal-rate Gaussian MAC", available online at arXiv:0908.1163v1 [cs.IT], August 2009.

\bibitem{ThC}
Thomas M Cover and J. A. Thomas, ''Elements of information theory'', \emph{second edition - Wiley Series in Telecommunications and Signal Processing}

\bibitem{RiU}
Rimoldi, B. and Urbanke, R, "A rate-splitting approach to the Gaussian multiple-access channel",  \emph{IEEE Transactions on Information Theory}, vol. 42, No. 02, pp. 364-375, 1996.
\end{thebibliography}
\end{document}